\def\L{{\cal L}}
\def\Lb{{\Lambda}}
\documentstyle[11pt]{article}

\def\appendix{\par
 \setcounter{section}{0}
 \setcounter{subsection}{0}
 \def\thesection{Appendix \Alph{section}}
 \def\theequation{\Alph{section}.\arabic{equation}}
 \setcounter{equation}{0}}

\begin{document}
\begin{flushright}
DO-TH 93-08 \\
SNUTP 91-12 \\
YUMS 93-05 \\
(March 1993)
\end{flushright}
\centerline{\Large \bf Flavor Democracy in Standard Models at High Energies}
\vspace{1.2cm}
\begin{center}
{{\bf G.~Cveti\v c} \\
Inst.~f\"ur Physik, Universit\"at Dortmund, 4600 Dortmund 50, Germany \\[0.8cm]
{\bf C.~S.~Kim} \\
Dept.~of Physics, Yonsei University, Seoul 120-749, Korea}
\end{center}
\vspace{1.2cm}
\centerline{\bf ABSTRACT}

It is possible that the standard model (SM) is replaced around some transition
energy $\Lb$ by a new, possibly Higgsless, ``flavor gauge theory''
such that the Yukawa (running) parameters of SM at $E \sim \Lb$ show up
an (approximate) flavor democracy (FD). We investigate the latter possibility
by studying the renormalization group equations for the Yukawa couplings
of SM with one and two Higgs doublets, by evolving them from given physical
values at low energies ($E \simeq 1 GeV$) to $\Lb$ ($ \sim \Lb_{pole}$)
and comparing the resulting fermion masses and CKM matrix elements at
$E \simeq \Lb$ for various $m_t^{phy}$ and
ratios $v_u/v_d$ of vacuum expectation values. We find
that the minimal SM and the closely related SM with two Higgs
doublets (type I) show increasing deviation from FD when energy is
increased, but that SM with two Higgs doublets (type II) clearly tends to FD
with increasing energy - in both the quark and the leptonic sector (q-q and
l-l FD). Furthermore, we
find within the type II model that, for $\Lb_{pole} \ll \Lb_{Planck}$,
$m_t^{phy}$ can be less than $200 GeV$ in most cases of chosen $v_u/v_d$.
Under the assumption that also the corresponding Yukawa couplings in
the quark and the leptonic sector at $E \simeq \Lb$ are equal (l-q FD),
we derive estimates of bounds on masses of top quark and tau-neutrino,
which are compatible with experimental bounds.

\newpage

\section{Introduction}
The hierarchical pattern of the quark masses, their mixing, and the
Higgs sector in general, remain outstanding issues of the
Standard model (SM) electroweak
theory. While a gauge interaction is characterized by its universal
coupling constant, the Yukawa interactions may have as many coupling
constants as there are pairs of fermionic fields coupled to neutral
Higgs bosons. There is no apparent underlying principle which governs
the hierarchy of the various Yukawa couplings. As a result, SM predicts
neither the fermion masses nor their mixing.

For this reason, we regard it as possible that around some high energy $\Lb$
($> 1 TeV$) SM breaks down and is replaced there by a new extended
gauge theory that is responsible for an (approximate) flavor democracy
of SM at high energies near the transition ($E \sim \Lb$). Flavor
democracy (FD) in a fermionic sector basically means equality of
Yukawa couplings in this sector in a flavor basis.

An extended
gauge theory responsible for FD we call ``flavor gauge theory'' (FGT).
In Appendix A we outline a possible simplified scheme for such an FGT, in
order to motivate the concept of flavor democracy. In Section 2 we
introduce formally the concept of FD for various fermionic sectors.
In Appendix B, possible parametrizations for the deviations from FD
are discussed.
In Section 3 we use 1-loop renormalization group equations (RGE's)
for Yukawa couplings of quarks in SM with one and two Higgs doublets,
by considering the second and the third families of quarks together.
We evolve RGE's from the low energy ($\sim 1 GeV$) physical mass scale
to high energy and investigate whether
the deviations from FD in the quark (up-type and down-type) sectors
decrease with increasing energy. We then investigate the same question
also in the leptonic sectors. Search for such trends toward FD can
be interpreted also as a somewhat unconventional search
(i.e., from the low energy physical scale to high energy)
within SM for signals of new physics.
In Section 4 we adopt the assumption of an additional lepton-quark FD
(i.e., that at high energies the Yukawa couplings in the corresponding
quark and leptonic sectors coincide at a high energy $E \simeq \Lb$), which
leads to estimates of bounds on masses of the top quark and tau-neutrino.
These turn out to be compatible with experimental bounds.
Section 5 contains further discussion of results. Conclusions of
this paper are summarized in Section 6.

\section{Flavor Democracy}
We define that SM at some energy $\mu \simeq \Lb$ possesses flavor democracy
(FD) if there exists a flavor basis \footnote{Flavor basis is a basis in
which the left-handed doublets of fermions transform as doublets under
$SU(2)_L$.} where the Yukawa coupling strengths are equal in certain
fermionic sectors. In this respect, we will restrict ourselves to two
cases of SM, as motivated in Appendix A: the minimal SM, and SM with two
Higgs doublets (type II)~\cite{dm}.

\vspace{0.4cm}

{\bf a) Minimal SM:}

FD would mean here an ``overall'' FD - with equal coupling strength for
the up-type and the down-type fermionic sector simultaneously (for quarks,
and separately for leptons):

\begin{eqnarray}
\L_{Yukawa}(\Lb) & \simeq &
         - G^{(q)} \sum_{i,j=1}^3 \lbrace
[( \bar q^{(i)}_L \tilde H ) q^{(j)}_{uR} + \mbox{h.c.}] +
[( \bar q^{(i)}_L        H ) q^{(j)}_{dR} + \mbox{h.c.}] \rbrace
\nonumber\\
& &      - G^{(l)} \sum_{i,j=1}^3 \lbrace
[( \bar l^{(i)}_L \tilde H ) l^{(j)}_{uR} + \mbox{h.c.}] +
[( \bar l^{(i)}_L        H ) l^{(j)}_{dR} + \mbox{h.c.}] \rbrace  \ ,
\end{eqnarray}
where we use the notation
\begin{displaymath}
H  =  {H^{+} \choose H^0} \ , \qquad \tilde H  = i \tau_2 H^{\ast} \ ,
\end{displaymath}
\begin{displaymath}
q^{(i)} = {q^{(i)}_u \choose q^{(i)}_d} \ , \qquad
q^{(1)} = {u \choose d} \ , \ q^{(2)} = {c \choose s} \ , \
q^{(3)} = {t \choose b} \ ,
\end{displaymath}
\begin{equation}
l^{(i)} = {l^{(i)}_u \choose l^{(i)}_d} \ , \qquad
l^{(1)} = {\nu_e \choose e} \ , \ l^{(2)} = {\nu_{\mu} \choose \mu} \ , \
l^{(3)} = {\nu_{\tau} \choose \tau} \ .
\label{eq:fermnotation}
\end{equation}

\vspace{0.5cm}

{\bf b) SM with two Higgs doublets $H^{(u)}$ and $H^{(d)}$ (type II):}

$H^{(u)}$ and $H^{(d)}$ couple in this SM to $f_{uR}$ and $f_{dR}$
fermion isosinglets, respectively~\cite{dm}.
FD would mean here an up-type FD and a separate down-type FD, in the quark
and the leptonic sector, i.~e. ``q-q up-type FD'', ``q-q down-type FD'',
``l-l up-type FD'' and ``l-l down-type FD''
\begin{eqnarray}
\L_{Yukawa}(\Lb) & \simeq &
- G_u^{(q)} \sum_{i,j=1}^3
[( \bar q^{(i)}_L \tilde H^{(u)}) q^{(j)}_{uR} + \mbox{h.c.}]
- G_d^{(q)} \sum_{i,j=1}^3
[( \bar q^{(i)}_L        H^{(d)}) q^{(j)}_{dR}+\mbox{h.c.}]
\nonumber\\
& &   - G_u^{(l)} \sum_{i,j=1}^3
[( \bar l^{(i)}_L \tilde H^{(u)}) l^{(j)}_{uR} + \mbox{h.c.}]
      - G_d^{(l)} \sum_{i,j=1}^3
[(\bar l^{(i)}_L         H^{(d)}) l^{(j)}_{dR} + \mbox{h.c.}] \ .
\label{eq:Higgs2SM}
\end{eqnarray}
For the mass matrices $M_u^{(q,l)}$ and $M_d^{(q,l)}$, FD implies
\begin{equation}
M_{\alpha}^{(q,l)} \simeq \frac{G_{\alpha}^{(q,l)} v_{\alpha}}{\sqrt{2}}
\left[ \begin{array}{ccc}
1 & 1 & 1 \\
1 & 1 & 1 \\
1 & 1 & 1
\end{array} \right] \ , \qquad (\alpha = u, d) \ ,
\label{eq:fd1}
\end{equation}
where $v_{\alpha} = \sqrt{2} \langle (H^{(\alpha)})^0 \rangle_o$  are
the vacuum expectation values (VEV).
Upon diagonalization, one obtains FD conditions for masses and mixings
at $\mu = \Lb$
\begin{equation}
M_{\alpha}^{(q,l)diag} \simeq \frac{3 G_{\alpha}^{(q,l)} v_{\alpha}}
{\sqrt{2}}
\left[ \begin{array}{ccc}
0 & 0 & 0 \\
0 & 0 & 0 \\
0 & 0 & 1
\end{array} \right] \ , \qquad
V_{ckm}^{(q,l)} \simeq
\left[ \begin{array}{ccc}
1 & 0 & 0 \\
0 & 1 & 0 \\
0 & 0 & 1
\end{array} \right] \ .
\label{eq:fd2}
\end{equation}
Note that even at low energies ($\mu \simeq 1 - 100 GeV$), the fermionic
mass spectrum is rather close to this form - i.e., the deviations from FD
in SM in the up-type, and the down-type sectors, are already at low
energies not large.

In our calculations, we will approximate the masses of the lightest quark
and leptonic family to be zero at energy $\mu_0 = 1 GeV$,
and will assume that their mixings to other families
($\propto (V_{ckm})_{u \beta}$, $\beta=s,b$) are not strong enough to affect
substantially the masses in those families at energies $\mu > \mu_0
(=1 GeV)$. These assumptions are based on previous investigations of
renormalization
group equations (RGE's) of SM (e.~g.~see~\cite{hp}). Furthermore, first
we will restrict our calculations to the quark sector ($(c,s)$ and $(t,b)$),
since the leptonic sector doesn't affect the evolution of the Yukawa
couplings for quarks substantially, as will be seen later. We adopt the
light quark masses (at $\mu_0 = 1 GeV$) obtained by Gasser and
Leutwyler~\cite{gl1}: $m_c=1.35 \pm 0.05 GeV$, $m_s=0.175 \pm 0.05 GeV$ and
$m_b=5.3 \pm 0.1 GeV$. For the (physical) mass of the top quark
$m_t^{phy} = m_t(\mu=m_t^{phy})$,
we consider the range between $100 GeV$ and $200 GeV$, to
explore the whole possible region of the experimentally predicted
$m_t^{phy}$~\cite{cdf,ef}. In this case of the two families~\footnote{Please
note that we are still considering SM with $n_f=3$, but the first
family being decoupled from the other two in the Yukawa sector.}, the
Cabibbo-Kobayashi-Maskawa matrix for quarks, at $\mu_0 = 1 GeV$,
is~\cite{gl2}
\begin{equation}
V_{ckm}^{(q)}(\mu=\mu_0) =
\left( \begin{array}{cc}
        1-\epsilon^{(q)}   & \eta^{(q)} \\
         - \eta^{(q)}      & 1-\epsilon^{(q)}
       \end{array} \right)
    \simeq
\left( \begin{array}{cc}
        1-1.25 \times 10^{-3} & 0.05 \\
       -0.05                  & 1 -1.25 \times 10^{-3}
       \end{array} \right) \ .
\end{equation}
The matrix is orthonormal, CP-violation effects are neglected.

The trend toward FD, as $\mu \uparrow \Lb$, would then mean, according to
(5) \footnote{Superscript $D$ in $m^D_{\nu}$ denotes that these are
Dirac neutrino masses; $\eta^{(q)} = (V_{ckm}^{(q)})_{cb}$, and $\eta^{(l)}$
is the corresponding element in the leptonic CKM matrix.}
\begin{displaymath}
\delta_u^{(q)} (=\frac{m_c}{m_t}) \ , \
\delta_u^{(l)} (=\frac{m^D_{\nu_{\mu}}}{m^D_{\nu_{\tau}}})
\longrightarrow 0 \ ,
\end{displaymath}
\begin{displaymath}
\delta_d^{(q)} (=\frac{m_s}{m_b}) \ , \
\delta_d^{(l)} (=\frac{m_{\mu}}{m_{\tau}}) \longrightarrow 0 \ ,
\end{displaymath}
\begin{equation}
\eta^{(q)} \ , \ \eta^{(l)} \longrightarrow 0 \ .
\label{eq:FD}
\end{equation}
In the case of the minimal SM (``overall FD'') we would have in addition
\begin{displaymath}
\frac{m_b}{m_t} \ , \ \frac{m_{\tau}}{m^D_{\nu_{\tau}}} \longrightarrow  1 \ .
\end{displaymath}
In Appendix B, we show how to find FD-flavor bases, i.~e.~ flavor bases
in which we have in the case of FD equal Yukawa coupling strengths
in fermionic sectors. There we argue that
in an ``up-down symmetric'' FD-flavor basis the parameters of deviation
from FD are
\begin{eqnarray}
\triangle_u^{(q)} & = & \sqrt{ \left( \frac{m_c}{m_t} \right)^2
+ \left( \frac{\eta^{(q)}}{2} \right)^2 } \ ,
\nonumber\\
\triangle_d^{(q)} & = & \sqrt{ \left( \frac{m_s}{m_b} \right)^2
+ \left( \frac{\eta^{(q)}}{2} \right)^2 } \ ,
\end{eqnarray}
and analogously in the leptonic sector. Hence, the above conditions
(eq. (5) ) are indeed equivalent to the trend to FD in the up-type
and the down-type sectors of quarks and leptons
\begin{displaymath}
\triangle_u^{(q,l)} \rightarrow 0 \qquad  \mbox{and} \qquad
\triangle_d^{(q,l)} \rightarrow 0 \ .
\end{displaymath}

\section{Analysis of Yukawa Couplings with RGE's in SM with one and two
Higgs Doublets}
\subsection{RGE's of SM for Yukawa Parameters}
The renormalization group equations (RGE's) for coupling parameters tell
us how these parameters effectively change when the energy $\mu$ of the
experimental probes changes. The 1-loop RGE's of SM for Yukawa parameters
have been systematically written down in literature (e.~g.~\cite{hlr} and
references therein), for different versions of SM.

In SM with two Higgs doublets (type II)~\cite{dm}, the Yukawa couplings
$U_{ij}^{(q)}$, $D_{ij}^{(q)}$,
$U_{ij}^{(l)}$ and $D_{ij}^{(l)}$ are defined as the couplings contained
in the following gauge invariant Lagrangian
\begin{eqnarray}
\L_{Yukawa} & = &          -  \sum_{i,j=1}^3 \lbrace
[( \bar q^{(i)}_L \tilde H^{(u)} ) q^{(j)}_{uR}U_{ij}^{(q)} + \mbox{h.c.}] +
[( \bar q^{(i)}_L        H^{(d)} ) q^{(j)}_{dR}D_{ij}^{(q)} + \mbox{h.c.}]
\rbrace
\nonumber\\
& &      -  \sum_{i,j=1}^3 \lbrace
[( \bar l^{(i)}_L \tilde H^{(u)} ) l^{(j)}_{uR}U_{ij}^{(l)} + \mbox{h.c.}] +
[( \bar l^{(i)}_L        H^{(d)} ) l^{(j)}_{dR}D_{ij}^{(l)} + \mbox{h.c.}]
\rbrace  \ .
\end{eqnarray}
Here we used the notations of eqs.~(2) and (3).
In the case of the minimal SM, we have to replace
$H^{(u)},H^{(d)} \mapsto H$ in the above form of $\L_{Yukawa}$.

We can deduce from~\cite{hlr} the corresponding 1-loop RGE's for the
Yukawa matrices
\begin{equation}
Q^{(u)} = U^{(q)}U^{(q)\dagger} \ , \ Q^{(d)} = D^{(q)}D^{(q)\dagger} \ , \
L^{(u)} = U^{(l)}U^{(l)\dagger} \ , \ L^{(d)} = D^{(l)}D^{(l)\dagger} \ .
\end{equation}
These RGE's are of the form
\begin{eqnarray}
32 \pi^2 \frac{d}{dt}Q^{(u)} & = & 3 Q^{(u)^2} +
 \kappa (Q^{(u)}Q^{(d)}+Q^{(d)}Q^{(u)}) + 2 Q^{(u)}(\Xi^{(q)}_u - A^{(q)}_u)
 \ , \nonumber\\
32 \pi^2 \frac{d}{dt}Q^{(d)} & = & 3 Q^{(d)^2} +
 \kappa (Q^{(u)}Q^{(d)}+Q^{(d)}Q^{(u)}) + 2 Q^{(d)}(\Xi^{(q)}_d - A^{(q)}_d)
 \ , \nonumber\\
32 \pi^2 \frac{d}{dt}L^{(u)} & = & 3 L^{(u)^2} +
 \kappa (L^{(u)}L^{(d)}+L^{(d)}L^{(u)}) + 2 L^{(u)}(\Xi^{(l)}_u - A^{(l)}_u)
 \ , \nonumber\\
32 \pi^2 \frac{d}{dt}L^{(d)} & = & 3 L^{(d)^2} +
 \kappa (L^{(u)}L^{(d)}+L^{(d)}L^{(u)}) + 2 L^{(d)}(\Xi^{(l)}_d - A^{(l)}_d)
 \ ,
\label{eq:RGE}
\end{eqnarray}
where
\begin{displaymath}
t = ln \left( \frac{2 \mu^2}{v^2} \right) \ ,
\end{displaymath}
\begin{displaymath}
A^{(q)}_u = \pi \left[ \frac{17}{3} \alpha_1+9 \alpha_2+32 \alpha_3 \right]
\ , \qquad A^{(q)}_d = A^{(q)}_u-4\pi\alpha_1 \ ,
\end{displaymath}
\begin{equation}
A^{(l)}_u = \pi [ 3\alpha_1+9\alpha_2] \ , \qquad
A^{(l)}_d = \pi [15\alpha_1+9\alpha_2] \ .
\end{equation}
Here, $\alpha_1$, $\alpha_2$ and $\alpha_3$ are the usual SM gauge
couplings corresponding to $U(1)_Y$, $SU(2)_L$ and $SU(3)_c$, respectively:
\begin{equation}
\alpha_i(\mu) = \alpha_i(M_Z)\left[ 1 + \frac{c_i}{4\pi}\alpha_i(M_Z)
ln\left( \frac{\mu^2}{M_Z^2} \right) \right]^{-1} \ ,
\end{equation}
with
\begin{displaymath}
  \alpha_1(M_Z) = 0.01013   \ , \qquad \alpha_2(M_Z) = 0.03322
 \ , \qquad \alpha_3(M_Z) = 0.109    \ .
\end{displaymath}
The other quantities in these equations differ from each other in the
two cases of SM

\vspace{0.3cm}

\begin{tabular}{|c|c|}
\hline
SM with two Higgs doublets (type II) & minimal SM \\
\hline
$\kappa = \frac{1}{2}$ & $\kappa = - \frac{3}{2}$ \\
$\Xi^{(q)}_u = \Xi^{(l)}_u = Tr(3Q^{(u)}+L^{(u)})$ &
$\Xi^{(q)}_{u,d}=\Xi^{(l)}_{u,d}=$ \\
$\Xi^{(q)}_d = \Xi^{(l)}_d = Tr(3Q^{(d)}+L^{(d)})$ &
$ = Tr(3Q^{(u)}+3Q^{(d)}+L^{(u)}+L^{(d)})$ \\
$c_1=-7 \ , \ c_2=3 \ , \ c_3=7$ & $c_1=-\frac{41}{6} \ , \ c_2=\frac{19}{6}
 \ , c_3=7$ \\
\hline
\end{tabular}

\vspace{0.3cm}

Note that the mass matrices are proportional to the VEV's of the Higgses
\begin{equation}
M_u^{(q,l)} = \frac{v_u}{\sqrt{2}}U^{(q,l)} \ , \qquad
M_d^{(q,l)} = \frac{v_d}{\sqrt{2}}D^{(q,l)} \ ,
\label{eq:mass}
\end{equation}
where
\begin{displaymath}
\langle H^{(u)} \rangle_o = \frac{1}{\sqrt{2}} {0 \choose v_u} \ , \qquad
\langle H^{(d)} \rangle_o = \frac{1}{\sqrt{2}} {0 \choose v_d} \ ,
\end{displaymath}
\begin{displaymath}
v_u^2+v_d^2=v^2 (=246^2 GeV^2) \ .
\end{displaymath}
For the minimal SM, we replace in the above relations $H^{(u)}, H^{(d)}
\mapsto H$ and $v_u,v_d \mapsto v$.
Hence, SM with two Higgs doublets (type II) has one additional, as yet
free, parameter $v_u/v_d$, which crucially influences the
running masses. Mathematically, we see this in eq.~(14) which implies
different boundary conditions in RGE's at $\mu = 1 GeV$ in comparison to
the minimal SM. Theoretical bounds on $v_u/v_d$ have been
investigated~\cite{bhp}, by requiring that certain processes involving
Higgs particles and top quarks are described well by perturbative approach.
This results in $v_u/v_d$ being between $0.1$ and $100$. On the other
hand, experimental evidence from $B-\bar B$ mixing, $D-\bar D$ mixing,
$\triangle m_K$, $\epsilon_K$ and missing $E_T$ measurements at $p \bar p$
colliders suggests more restrictive bounds~\cite{bhp}
\begin{equation}
0.3 \ < \ \frac{v_u}{v_d} \ < \ 10 \ .
\end{equation}

\subsection{Numerical Results and Discussions - Quark Sector Only}

First we limit ourselves to the quark sector - $2 \times 2$ matrices
corresponding to the two families (c,s) and (t,b),
taking the leptonic $2 \times 2$ matrices in
the RGE's~(11) to be zero. In this case, the low energy boundary
conditions for the masses of light quarks and the mixing are well
known~\cite{gl1,gl2}, and $m_t^{phy} \sim 100-200 GeV$. Fig.~1 shows
the running of quark masses for a typical value of VEV ratio $v_u/v_d=1$.
The running of the flavor democracy parameters $\delta_u^{(q)}(=m_c/m_t)$,
$\delta_d^{(q)}(=m_s/m_b)$ and $\eta^{(q)}(=(V_{ckm}^{(q)})_{cb})$ in this
case, and in the case of
$v_u/v_d=0.5$, are depicted in Figs.~2,~3, respectively. These
quantities for the minimal SM are depicted in Fig.~4. The ratio of the
Yukawa couplings $(g_b/g_t)$ as function of energy, for various ratios
of VEV's, is shown in Fig.~5.

In the minimal SM, quark masses reach their poles $(m_q \to \infty)$
before the Planck scale only for very heavy top ($m_t^{phy}\geq215 GeV$),
as can be seen also from Fig.~4 (for $m_t^{phy}=250 GeV \Rightarrow
\Lambda_{pole} \approx 10^{10.3}GeV$). On the other hand, the FD
parameters decrease (i.~e.~positive trend to FD) only in the up-type
sector: $\delta_u^{(q)}$ decreases, while the parameter
$\delta_d^{(q)}$ increases always with energy, as well as does
the mixing parameter $\eta^{(q)}$. Furthermore, $m_b/m_t \to 0$,
rather than $m_b/m_t \to 1$. The conditions for FD (eq.~(7))
are clearly not fulfilled in the minimal SM.

Results for SM with two Higgs doublets (type II), depicted in
Figs.~1-3, show a more interesting structure. In all these cases,
all the FD parameters ($\delta_u^{(q)}$, $\delta_d^{(q)}$ and $\eta^{(q)}$)
are decreasing functions of energy. Hence, here we have a clear trend
to FD (7). Furthermore, in most of these cases, the poles
$\Lambda_{pole}$ for the masses are reached long before $\Lambda_{Planck}$:
$\log\Lambda_{pole} < \log\Lambda_{Planck}$. Table 1 contains
$\Lambda_{pole}$ and $\Lambda_{pert}$ for various $v_u/v_d$ and
$m_t^{phy}$. $\Lambda_{pert}$ is
the energy where the perturbative approach mathematically breaks down
\footnote{
It was determined by requiring that the 2-loop correction to $dg_t/dt$ in
RGE's for Yukawa couplings in the minimal SM~\cite{mv}
be approximately equal at $\Lambda_{pert}$ to the corresponding 1-loop
contribution, resulting in $g_t(\mu=\Lambda_{pert})=\sqrt{6}\pi$. We use
this relation also as a reasonable estimate in SM with two Higgs doublets.}.
We can interpret these quantities as being of the order of magnitude of the
transition energy $\Lambda_{trans}$ between SM and some extended
flavor gauge theory (FGT): $\log\Lambda_{trans} \simeq
\log\Lambda_{pert} \simeq \log\Lambda_{pole}$. The clear trend to FD
at these energies {\it within} SM could be interpreted as a signal
of a corresponding new physics.

\subsection{Results and Discussions - Leptonic Sector Included}

By including also the leptonic sector in the numerical investigations of
the RGE's~(11), we were able to investigate also the behavior of
the corresponding FD parameters separately in the leptonic sector:
$\delta_u^{(l)}(=m^D_{\nu_{\mu}}/m^D_{\nu_{\tau}})$,
$\delta_d^{(l)}(=m_{\mu}/m_{\tau})$ and
$\eta^{(l)}(=(V_{ckm}^{(l)})_{\nu_{\mu}\tau})$.

However, for the boundary conditions of RGE's at low energy ($\mu = 1 GeV$)
we need the masses of $\mu$, $\tau$, $\nu_{\mu}^{Dirac}$ and
$\nu_{\tau}^{Dirac}$. While the masses of $\mu$ and $\tau$ are well
known ($m_{\mu}^{phy}=0.106 GeV$, $m_{\tau}^{phy}=1.78 GeV$),
the Dirac neutrino masses are not. However, by invoking the
usual ``see-saw mechanism''~\cite{grs}
and assuming that $\Lb_{trans}$($\sim
\Lambda_{pole}$) is of the order of magnitude of the Majorana mass $M_R$,
we can estimate the masses of Dirac neutrinos at low energy
(see Section 4 and eq.~(19) for more details)
\begin{equation}
m^{D}_{\nu} \sim \sqrt{M_R m^{phy}_{\nu}} \sim
 \sqrt{\Lambda_{pole} m^{phy}_{\nu}} \ .
\end{equation}
Using the upper bounds on $m^{phys}_{\nu_{\mu}}$($<250 keV$) and
$m^{phys}_{\nu_{\tau}}$($<31 MeV$) as suggested by experiment~\cite{argus},
we obtain for a
typical case of $v_u/v_d=1$ and $m_t^{phy}=200 GeV$ the low energy
Dirac neutrino masses $m^D_{\nu_{\tau}} \sim 100 GeV$ and
$m^D_{\nu_{\mu}} \sim 10 GeV$. We use these estimated values in the
boundary conditions for the RGE's~(11) to investigate the
FD behavior simultaneously for the quark and the leptonic sector.
The FD behavior in the leptonic sector turns out to be completely
analogous to that in the quark sector: in the minimal SM $\delta_d^{(l)}$
and $\eta^{(l)}$ increase, while in SM with two Higgs doublets (type II)
all leptonic FD parameters ($\delta_u^{(l)}$, $\delta_d^{(l)}$ and
$\eta^{(l)}$) decrease with increasing energy and we have a clear trend
to FD also in the leptonic sector. In Table 2 we display some of the
pertaining results. It is to be noted that the transition energy
($\simeq \Lambda_{pole}$) is in most cases diminished on the logarithmic
scale by about 5-10 per cent with the inclusion of the leptonic sector.
Furthermore, $m_b/m_t$ and $m_{\tau}/m^D_{\nu_{\tau}}$
decrease with increasing energy.

We have numerically checked also that these conclusions are
qualitatively independent of the specific chosen Dirac neutrino masses at
low energy $m^D_{\nu}(\mu=1 GeV)$.

\section{Top and Neutrino Mass Estimates
         - under the Assumption of Quark-Lepton Flavor Democracy}

In this section we extend our investigation by imposing, {\it in addition}
to the conditions (7) for FD in the quark (q-q) and leptonic (l-l) sector,
the condition of FD for the (combined) quark-leptonic (q-l) sector
\begin{equation}
 \frac{m^D_{\nu_{\tau}}}{m_t} \ , \ \frac{m_{\tau}}{m_b} \rightarrow 1
\qquad \mbox{as} \ E \rightarrow \Lambda_{trans} \ .
\label{eq:FDN1}
\end{equation}
Note that this would correspond, in the FGT-scheme of Appendix A, to the
assumption that FGT treats quarks and leptons on equal grounds:
$\kappa_{\alpha}^{(q)} = \kappa_{\alpha}^{(l)}$, (for $\alpha = u,d$) in
eq.~(A.4). In eq.~(3), this would correspond to
$G_{\alpha}^{(q)}=G_{\alpha}^{(l)}$ (for $\alpha=u,d$).

In the calculation here, we take into account only the third families
$(t,b)$ and $(\nu^D_{\tau}, \tau)$, and investigate the running of
their Yukawa couplings (or equivalently: their masses) with increasing
energy. It can be checked that the lighter second families do not affect
the behavior of the third families in any appreciable way (vice-versa is
not true). Here, we approximate, for simplicity, the transition energy
$\Lambda_{trans}$ between SM and an extended theory (e.~g.~FGT) with
$\Lambda_{pole}$. Such an approximation may look reasonable in
view of the fact that we have only the indications for the relation
$\log\Lambda_{trans} \simeq \log\Lambda_{pole}$. Hence, our
additional (high energy) boundary conditions for the RGE's will be
approximated as
\begin{equation}
 \frac{m^D_{\nu_{\tau}}}{m_t}=1 \ , \qquad
 \frac{m_{\tau}}{m_b} = 1
\qquad \mbox{at} \ E \approx \Lambda_{pole} \ .
\label{eq:FDN2}
\end{equation}
These conditions are taken into account in our numerical calculations,
together with the known low energy boundary conditions
$m^{phy}_{\tau}=1.78 GeV$, $m_b(\mu=1 GeV)=5.3 GeV$,
$m^{phy}_t=100,~150,~200,~ 250 GeV$.
For chosen values of the VEV ratio $v_u/v_d$ and $m_t$, we found the masses
of Dirac tau-neutrino $m^D_{\nu_{\tau}}$ at $\mu=1 GeV$ which satisfy
the above boundary
conditions, by using numerical integration of RGE's from $\mu=1 GeV$ to
$\Lambda_{pole}$. For illustration, the resulting curves for one specific
case ($v_u/v_d = 1$ and $m_t^{phy} = 200 GeV$) are depicted in Fig.~6. The
results for $\Lambda_{trans}$ (taken as equal to $\Lambda_{pole}$) and
$m^D_{\nu_{\tau}}$ (at $\mu=1 GeV$),
for various values of $v_u/v_d$ and $m_t^{phy}$, are given
in Table 3. Since the Dirac neutrino masses shown in Table 3 are too large
to be compatible with results of the available experimental
predictions~\cite{argus}, we have to invoke the usual ``see-saw''
mechanism~\cite{grs} of the mixing of the Dirac neutrino masses (at
low energy) and the
much larger Majorana neutrino masses $M_R$, in order to derive the small
physical neutrino mass
\begin{equation}
m^{phy}_{\nu} \approx \frac{(m^D_{\nu})^2}{M_R} \ .
\end{equation}
Since the Majorana mass term breaks the lepton number conservation, the
Majorana masses are expected to be of the order of some new (unification)
scale $\Lambda$ ($\gg {\cal O}(M_W)$), and are usually assumed to be
$M_R \simeq \Lambda$. Within our context, the simplest choice of this
new unification scale would be the energy $\Lambda_{trans}$ where
SM is replaced by an extended gauge theory (e.~g.~an FGT). Since we
took here $\Lambda_{trans} \approx \Lambda_{pole}$, we obtain
\begin{equation}
m^{phy}_{\nu} \approx \frac{(m^D_{\nu})^2}{\Lambda_{pole}} \ .
\end{equation}
The last entry in Table 3 is calculated according to this equation.
As we can see, the physical tau-neutrino masses $m^{phy}_{\nu_{\tau}}$
predicted in this way are very small for the most cases of the choice
$v_u/v_d$ and $m^{phy}_t$, i.~e.~in most cases they are below the
experimentally predicted upper bounds~\cite{argus}.

Note that the scenario in this section, leading to our predictions
of $m^{phy}_{\nu_{\tau}}$, implicitly contains the following assumptions
that are not contained in the rest of the paper:

(a) An extended
flavor democratic gauge theory (FGT) treats the quark and the leptonic
sectors ``on equal grounds'', i.~e.~implies the equality of the
corresponding quark and leptonic Yukawa couplings near the FGT-SM transition
energy $\Lambda_{trans}$; (b) FGT contains in addition Majorana
neutrinos, and its energy range of validity also provides the scale for
the heavy Majorana masses (i.~e.~ $M_R \sim \Lambda_{trans}$);
(c) At energies of validity of SM (with two Higgs doublets, type II),
we consider that Majorana neutrinos remain decoupled
(or very weakly coupled) to the Dirac neutrinos.

In general, we could assume ${\cal O} (M_R) \simeq {\cal O}
( \Lambda_{new scale}) \geq
{\cal O}(\Lambda_{pole})$, and we would consequently end up with
even smaller $m^{phy}_{\nu}$ than those obtained in Table 3.

As seen in Table 3, when increasing $m^{phy}_t$ at a fixed $v_u/v_d$,
$m^D_{\nu_{\tau}}$ at $\mu = 1 GeV$
increases and $\Lambda_{pole}$ decreases, hence
$m^{phy}_{\nu_{\tau}}$ increases. This provides us with upper and
lower bounds on the values of $m_t^{phy}$ (at a given $v_u/v_d$) for
various specific upper bounds imposed on $m^{phy}_{\nu_{\tau}}$
(e.~g.~ $\leq 31 MeV$~\cite{argus}, or $ \leq 1MeV$, or
$ \leq 17 keV$~\cite{hpr}) and on $\Lambda_{pole}$ (e.~g.~$ \leq
\Lambda_{Planck}$, or $ \leq 10^{10} GeV$, or $ \leq 10^5 GeV$),
respectively. These bounds on $m^{phy}_t$ are depicted as functions
of $v_u/v_d$ in Fig.~7. Even with the largest possible upper bounds on
$m^{phy}_{\nu_{\tau}}$ ($\leq 31 MeV$) and $\Lambda_{pole}$ ($ \leq
\Lambda_{Planck}$), we can still get the narrow bands on the values of
$m^{phy}_t$ at any given $v_u/v_d$. For example, if $v_u/v_d = 1$, then
$m_t^{phy}$ ranges between $150 GeV$ and $215 GeV$. Inversely, if
$m^{phy}_t = 150 GeV$, then $0.5 \leq v_u/v_d \leq 1.0$.

\section{Further Discussions}
The calculations in the paper, except for the previous section,
were carried out by including the CKM mixing between the considered
second and third families of quarks, by taking
$(V_{ckm}^{(q)})_{cb} (\mu=1GeV) =0.05$,
i.e.~the experimentally suggested value for the quark
sector~\cite{gl2}. For simplicity, we took $(V_{ckm}^{(l)})_{\nu_{\mu}\tau}
(\mu=1GeV) =0.05$, the same value as in the quark sector.
If we ignored any of the flavor mixings (this would
imply that the corresponding CKM matrix is the $2 \times 2$ identity matrix
at any energy),
the results for the masses and the FD parameters $\delta_u^{(q,l)}$ and
$\delta_d^{(q,l)}$ would be minimally changed - in general for about
1 per cent
or less, as numerically checked. Furthermore, as argued in
section 2, the lightest fermion families $(u,d)$ and $(\nu^D_e,e)$,
being practically massless, do not affect the evolution behavior of
the other families in any appreciable way (vice-versa is not true)
and hence do not affect the conclusions of this paper.

We have not dealt with another possibility of the minimally extended
SM's - the ones with two Higgs doublets of which only one contributes
to the quark masses (two Higgs doublets, type I~\cite{dm,hlr}).
Such extensions turn
out to have basically equal RGE's in the quark sector as
the minimal SM.
However, we have in such a case different quark mass relations than in
the minimal SM: $g_q(\mu)=\sqrt{2}m_q(\mu)/v_1$, where $v_1$ is VEV of
the Higgs coupled to the quarks, $v_1 < v=\sqrt{v_1^2+v_2^2}$. This would
imply that for the same $m_t^{phy}$, the low energy Yukawa coupling
$g_t(\mu=1 GeV)$ is larger in this model than in the minimal SM. Hence,
$\Lambda_{pole}$ ($<\Lambda_{Planck}$) would exist for a certain range
$m_t^{phy} \geq (v_1/v) \times (215 GeV)$, a range including values well below
$200 GeV$, while in the case of the minimal SM $m_t^{phy} \geq 215 GeV$.
However, the running of the other Yukawa couplings (including the
leptonic) is, as a rule, very much dominated by the running of $g_t$
(and not vice-versa), as in the minimal SM. Among other things, this
would imply that $\delta_d^{(q)}$ and $\eta^{(q)}$ would again, as in
the minimal SM, always increase (away from FD) with increasing energy,
as in Fig.~4.

In this paper, we did not consider the Higgs masses and their evolution
with energy. We were allowed to ignore them because, unlike the case
of the 2-loop RGE's, the 1-loop RGE's for Yukawa couplings do not
couple to parameters of the Higgs potential.

The results of the paper, with the view to flavor democracy,
advocate a specific extension of the minimal SM
- that with two Higgs doublets (type II), where one Higgs doublet couples
primarily to up-type fermions ($c, t, \nu^D_{\mu}$, etc.),
and the other to the down-type fermions ($s, b, \mu$, etc.). Note that
this extended SM automatically satisfies the relation $\rho \equiv
M_W^2/M_Z^2 \cos^2\theta_w = 1$ and has no FCNC at the tree level. Motivated
by the results of this paper, we express the conjecture that SM with two
Higgs doublets (type II), when naturally connected at $E=
\Lambda_{trans}$ with an extended gauge theory (FGT, see also
Appendix A), may yield a realistic mechanism for condensation of
two Higgses, $\langle \bar t t + \ldots \rangle$ and
$\langle \bar b  b + \ldots \rangle$~\cite{h2condens}.

Let us emphasize that applying RGE's of SM from low to high energies
(and not vice-versa), together with the notion of flavor democracy and
the related parameters $\delta_u^{(\alpha)}$, $\delta_d^{(\alpha)}$ and
$\eta^{(\alpha)}$ ($\alpha = q, l$), was crucial to get {\it within}
SM some signals of possible new physics. With the exception of the
section on the physical neutrino masses, the approach of this paper
can also be regarded as conservative and independent of any specific
assumptions.

There remain many possibilities to extend and continue the work presented
in this paper. We have already mentioned the possible investigation of
the running Higgs masses, as well as the possible condensation mechanism
for the (two) Higgs doublets which could lead to FD.
Other investigations could concentrate on the possibility that an extension
of SM (e.~g.~SUSY, fourth-generation, etc.~) could become effective in an
intermediate energy region between SM- and a new FGT-energy region.
Furthermore, several FGT schemes (with various choices of symmetry groups)
could be investigated for the regions $E>\Lambda_{trans}$.

\section{Conclusion}
We investigated the behavior of Yukawa couplings of SM in order to see
whether they show trends to flavor democratic (FD) structures as the
energy of the probes increases. We found that the minimal SM (and the
closely related SM with two Higgs doublets, type I) does not show such
trends in the sector of the down-type fermions ($s,b,\mu,\tau$) and
in the CKM-mixing behavior. On the other hand, we found that SM with
two Higgs doublets (type II) clearly shows such trends in all sectors
(down-type and up-type fermions), as well as in the CKM-mixing behavior.
These results may represent some signals of new physics
(``flavor democratic'') beyond SM. Then SM with two Higgs doublets
(type II) would be the preferred choice, and $m_t^{phy}$ would be
less than $200 GeV$ (for $\Lambda_{trans} \ll \Lambda_{Planck}$)
for most cases of chosen VEV ratio $v_u/v_d$.

Under the assumption that the corresponding Yukawa couplings
in the quark and the leptonic sectors in this model are equal
at $E \simeq \Lambda_{trans}$ (l-q FD), we were able to
estimate the masses of top quark and tau-neutrino.

\vspace{1.cm}

\section{Acknowledgement}
We would like to thank K.J.~Abraham, R.~B\"onisch, H.~Lange
and E.A.~Paschos for helpful discussions.
G.C.~wishes to thank the Deutsche Forschungsgemeinschaft (DFG) and
Dortmund University for financial support during the progress of this work.
C.S.K.~wishes to
thank the German Bundesministerium f\"ur Forschung und Technologie (BMFT)
for financial support during part of the progress of this work. The work
of C.S.K.~was also supported in part by the Korean Science and Engineering
Foundation, in part by the Center for Theoretical Physics, Seoul National
University, and in part by Yonsei University Faculty Research Grant.

\newpage
\begin{appendix}
\section[]{Simple Scheme of a ``Flavor Gauge Theory''}
\setcounter{equation}{0}

Here we briefly introduce a simple scheme for a ``flavor gauge theory''
(FGT) and point out its relation to the Yukawa sector of SM. The
fermionic sector of such an FGT Lagrangian that contains new gauge
boson $B_{\mu}$ and gauge coupling $g$ is written schmatically as
\begin{equation}
\L^{(FGT)}_{G-f} = - g \bar \Psi \gamma^{\mu} B_{\mu} \Psi \qquad
(\mbox{for} \ E \geq \Lambda) \ ,
\end{equation}
where $\Psi$ is the column of all fermions $f_j$ (in a usual flavor
basis of SM), and
$B_{\mu}=B^j_{\mu}T_j$~, with $T_j$ being the generator matrices of the
(new) symmetry group $G$. For simplicity, we denote here the transition
energy $\Lambda_{trans}$ between FGT and SM as $\Lambda$. We omit
the color indices of quarks. The $T_j$'s corresponding to the electrically
neutral $B^j_{\mu}$ are taken to be proportional to identity matrices in
the flavor space, i.~e.~FGT does not ``see'' flavors. The effective
neutral current-current interaction at ``low''-energy $E$ ($\Lambda \
\leq E \ll M_B$, $M_B$ being the mass of $B_{\mu}$) can then be written as
\begin{equation}
\L^{(FGT)}_{4f} \approx - \frac{g^2}{2M_B^2} \sum_{i,j}
(\bar f_i \gamma^{\mu} f_i)(\bar f_j \gamma_{\mu} f_j) \ .
\end{equation}
The left-to-right parts of the quark and lepton terms in the above sum
are those
which can contribute to Yukawa interactions. These parts can be
re-expressed, following a suggestion by B\"onisch~\cite{bo},
by applying a Fierz transformation~\cite{svz}
\begin{equation}
\L^{(FGT)}_{L-R} \approx \frac{2g^2}{M_B^2} \sum_{i,j}
\left[ (\bar q_{iL} q_{jR})(\bar q_{jR} q_{iL}) +
       (\bar l_{iL} l_{jR})(\bar l_{jR} l_{iL}) \right]
\qquad
(\mbox{for} \ \Lambda < E < M_B ) \ .
\label{eq:Fierz}
\end{equation}

We introduce two auxiliary scalar doublets $H^{(u)}$ and $H^{(d)}$
and add to the above Lagrangian the following squares (this
transformation doesn't change the path integral for the generating
functional, and hence formally doesn't change the physics~\cite{eg})
\begin{eqnarray}
\triangle \L & = &
- \sum_{f=q,l} \sum_{i,j=1}^3
      [ \kappa_u^{(f)} \tilde H^{(u) \dagger} +
        \frac{\sqrt{2}g}{M_B} \bar f^{(i)}_L f^{(j)}_{uR} ]
      [ \kappa_u^{(f)} \tilde H^{(u)} +
        \frac{\sqrt{2}g}{M_B} \bar f^{(j)}_{uR} f^{(i)}_L ]
\nonumber\\
& & - \sum_{f=q,l} \sum_{i,j=1}^3
      [ \kappa_d^{(f)} H^{(d) \dagger} +
        \frac{\sqrt{2}g}{M_B} \bar f^{(i)}_L f^{(j)}_{dR} ]
      [ \kappa_d^{(f)} H^{(d)} +
        \frac{\sqrt{2}g}{M_B} \bar f^{(j)}_{dR} f^{(i)}_L ] \ ,
\end{eqnarray}
where we used the notations of eq.~(2), and $\kappa_{\alpha}^{(q,l)}$ are
unspecified masses. Then the Lagrangian (A.3)
acquires the form of $\L_{Yukawa}(\Lb)$ of eq.~(3) (with
$G_{\alpha}^{(q,l)}= \kappa_{\alpha}^{(q,l)}g\sqrt{2}/M_B$) plus bare mass
terms of Higgs doublets
\begin{equation}
\L = (\L^{(FGT)}_{L-R}+ \triangle \L) =
\L_{Yukawa}(\Lb) - \sum_{\alpha=u,d}
(\kappa_{\alpha}^{(q)2}+\kappa_{\alpha}^{(l)2})
 (H^{(\alpha)\dagger}H^{(\alpha)}) \ .
\end{equation}
Several authors (\cite{h2condens} and references therein)
argued that the auxiliary Higgses in a class of Lagrangians,
including this Lagrangian,
can become dynamical by quantum effects, becoming
basically condensates of the $<\bar f f>$-type, when the energy
of the probes $E$ is decreased below $\Lb$ (=$\Lb_{trans}$) into
the SM energy region. The Higgs doublets $H^{(\alpha)}$ then become
renormalized by a normalization constant $Z_{\alpha}(E)$ ( which is
increasing as $E \uparrow \Lambda$), and we arrive at the flavor democratic
Yukawa Lagrangian of eqs.~(3)-(5) at $E \simeq \Lambda$, with the following
substitution
\begin{equation}
H^{(\alpha)} \mapsto \frac{H^{(\alpha)}}{Z^{1/2}_{\alpha}(E)} \
\Rightarrow \
G_{\alpha}^{(q,l)} =
\frac{\kappa_{\alpha}^{(q,l)} g \sqrt{2}}{M_BZ^{1/2}_{\alpha}(E)} \ ,
\qquad (\alpha = u,d) \ .
\end{equation}
In such a case, we would end up at low energies with the SM
with two Higgs doublets
(type II) which tends to flavor democracy (FD) as the energy of the
probes increases to the transition energy $\Lambda$.
If we expected transition of FGT into the minimal SM, we would have to
replace in the above relations
\begin{equation}
H^{(u)}, \ H^{(d)} \mapsto H \, \qquad
\kappa_u^{(q,l)}, \ \kappa_d^{(q,l)} \mapsto \kappa^{(q,l)} \ ,
\end{equation}
which would lead us to the flavor democratic Yukawa interaction of
eq.~(1).

We stress that the presented scenario is just a rather primitive scheme
of an FGT which may motivate the assumption of the trend to FD in SM
as we increase the energy of the probes. There remain many interesting
problems connected with such scenarios. One question is how
the condensation mechanism works precisely in this case.
There is also a question
of how to avoid the creation of too many physical Goldstones resulting
from the breaking of the horizontal global symmetry $SU(3)_L \times SU(3)_R$
(-the group of unitary transformations between the three
families)~\cite{zb}.

\section[]{Parameters for Deviations from Flavor Democracy}
\setcounter{equation}{0}
We define as an FD-flavor basis the flavor basis in which the Yukawa
coupling strengths become equal (separately in the up-type
and in the down-type sector, see eq.~(4)),
if the mass spectrum and the mixings are of the form of eq.~(5).
For simplicity, we will here
omit the subscripts and superscripts $(q,l)$ for quarks and leptons,
and restrict ourselves to the
case of two families ($2 \times 2$ matrices) in a specific (either $q$
or $l$) sector.

Transformation of the Yukawa matrices for the up-type and the down-type
sectors from the mass basis to a general flavor basis is of the form
\begin{equation}
U=V_L^{u\dagger}U_{diag}V_R^u \ , \qquad D = V_L^{d\dagger}D_{diag}V_R^d \ .
\end{equation}
$U_{diag}$ and $D_{diag}$ are the mass matrices divided by the
corresponding VEV's ($v/\sqrt{2}$ in the case of the minimal SM, and
$v_{u,d}/\sqrt{2}$ in the case of SM with two Higgs doublets, type II).
$V_{L,R}^u$ and $V_{L,R}^d$ are orthonormal matrices (the case of
no CP-violation assumed) satisfying the
relation
\begin{equation}
V_L^uV_L^{d\dagger}= V_{ckm} \ .
\end{equation}

It is straightforward to see that we have a large set of FD-flavor bases
\begin{equation}
U_{FD}^{(\theta)} = O^{\dagger} R(-\theta) U_{diag} O \ , \qquad
D_{FD}^{(\theta)} = O^{\dagger} R(-\theta) V_{ckm}D_{diag} O =
       O^{\dagger}R(\theta_{ckm}-\theta) D_{diag} O \ ,
\end{equation}
where we denoted
\begin{displaymath}
O = \frac{1}{\sqrt{2}}
\left( \begin{array}{cr}
       1 & -1 \\
       1 & 1
       \end{array} \right) \ , \qquad
R(-\theta) = \left(\begin{array}{cr}
                   \cos\theta & -\sin\theta \\
                   \sin\theta & \cos\theta
                   \end{array} \right)  \qquad
(0 \leq \theta \leq \theta_{ckm}) \ ,
\end{displaymath}
and
\begin{equation}
V_{ckm}= \left( \begin{array}{cc}
                (1-\epsilon) & \eta \\
                 -\eta       & (1-\epsilon)
                \end{array} \right) = R(\theta_{ckm}) \ .
\end{equation}
For each $\theta$ between zero and $\theta_{ckm}$ ($=\arcsin\eta$)
we have an FD-flavor basis $FD{(\theta)}$.
Taking the usual metrics for the matrices
\begin{equation}
\parallel A \parallel = \parallel \{ a_{i,j} \} \parallel =
\sqrt{\sum a_{ij}^2} \ \ ,
\end{equation}
we can define the deviations from FD for the up-type sector in such a basis
as
\begin{equation}
\triangle_{u,\theta} = \parallel \frac{U_{FD}^{(\theta)}}
{\parallel U_{FD}^{(\theta)} \parallel } - \frac{1}{2}
\left( \begin{array}{cc}
       1 & 1 \\
       1 & 1
       \end{array} \right) \parallel \ ,
\end{equation}
and analogously for the down-type sector. Explicit calculations then yield
\begin{displaymath}
\triangle_{u,\theta}  =
\sqrt{2} \left[ 1 - \frac{\cos{\theta}}{\sqrt{1+\delta_u^2}} \right]^{1/2} =
\sqrt{\delta_u^2 + \theta^2}
\left[ 1 + {\cal O}(\delta_u^2,\theta_{ckm}^2) \right] \ ,
\end{displaymath}
\begin{equation}
\triangle_{d,\theta}   =
\sqrt{2} \left[ 1 - \frac{\cos{(\theta_{ckm}- \theta)}}
{\sqrt{1+\delta_d^2}} \right]^{1/2}  =
\sqrt{\delta_d^2 + (\theta_{ckm}-\theta)^2}
\left[ 1 + {\cal O}(\delta_d^2,\theta_{ckm}^2) \right] \ ,
\end{equation}
where we denoted $\delta_u=m_c/m_t$, $\delta_d = m_s/m_b$ for the quark
sector, and analogously for the leptonic sector.

Choosing the ``up-down symmetric'' FD-flavor basis ($\theta=\theta_{ckm}/2
\simeq \eta/2$), we obtain
\begin{equation}
\triangle_u \simeq \sqrt{\delta_u^2 + \frac{\eta^2}{4}} \ , \qquad
\triangle_d \simeq \sqrt{\delta_d^2 + \frac{\eta^2}{4}} \ .
\end{equation}
This apears to be the most reasonable choice to parametrize deviations
from FD in the up-type and the down-type sectors (for quarks, and for
leptons). On the other hand, various choices of $\theta$ ($0 \leq
\theta \leq \theta_{ckm}$) would give us similar values for deviations.
All these parameters are, of course, functions of the running energy.
\end{appendix}

\newpage

\noindent {\Large\bf Tables}

\vspace{1.cm}
\begin{table}[h]
 \begin{center}
 Table 1\\
\vspace{0.5cm}
  \begin{tabular}{|c|c|l|l|}  \hline
$v_u/v_d$ & $m_t^{phy}$ & $\Lambda_{pert}$ &  $\Lambda_{pole}$ \\ \hline
          &  $100$      & $(3.16)\cdot10^{17}$  &  $(4.26)\cdot10^{17} $ \\
\cline{2-4}
  $0.5$   &  $150$      & $(1.20)\cdot10^5$     &  $(1.61)\cdot10^5$     \\
\cline{2-4}
          &  $200$      & $(3.80)\cdot10^3$     &  $(5.11)\cdot10^3$     \\
\cline{2-4}
          &  $250$      & $(1.20)\cdot10^3$     &  $(1.70)\cdot10^3$     \\
\hline
          &  $100$   &$>\Lambda_{Planck}$&$>\Lambda_{Planck}$  \\ \cline{2-4}
  $1.0$   &  $150$   &$>\Lambda_{Planck}$&$>\Lambda_{Planck}$  \\ \cline{2-4}
          &  $200$      & $(1.84)\cdot10^7$     &   $(2.48)\cdot10^7$    \\
\cline{2-4}
          &  $250$      & $(6.73)\cdot10^4$     &   $(9.08)\cdot10^4$    \\
\hline
          &  $100$   &$>\Lambda_{Planck}$&$>\Lambda_{Planck}$  \\ \cline{2-4}
  $5.0$   &  $150$   &$>\Lambda_{Planck}$&$>\Lambda_{Planck}$  \\ \cline{2-4}
          &  $200$   &$>\Lambda_{Planck}$&$>\Lambda_{Planck}$  \\ \cline{2-4}
          &  $250$      & $(3.88)\cdot10^9$     &   $(5.22)\cdot10^9$    \\
\hline
          &  $100$   &$>\Lambda_{Planck}$&$>\Lambda_{Planck}$  \\ \cline{2-4}
min.~SM   &  $150$   &$>\Lambda_{Planck}$&$>\Lambda_{Planck}$  \\ \cline{2-4}
          &  $200$   &$>\Lambda_{Planck}$&$>\Lambda_{Planck}$  \\ \cline{2-4}
          &  $250$      & $(1.5)\cdot10^{10}$   &   $(2.0)\cdot10^{10}$   \\
\hline
  \end{tabular}
\vspace{0.5cm}
\caption{$\Lb_{pole}$ and $\Lb_{pert}$ for various $m_t^{phy}$
and for various VEV ratios $v_u/v_d$ (in SM with two Higgs doublets,
type II) and for the the minimal SM. $\Lb_{pole}$ is the energy where
the quark masses become infinite, $\Lb_{pert}$ the energy where the
perturbative approach breaks down. The masses of leptons were taken to
be zero. All energies are in $GeV$. $\Lb_{Planck} \simeq 10^{19} GeV$.}
 \end{center}
\end{table}

\begin{table}[h]
\vspace{0.4cm}
 \begin{center}
Table 2\\
\vspace{0.5cm}
  \begin{tabular}{|c|c|c||l|l||l|l|l|l|l|}  \hline
$v_u/v_d$  & $m_t^{phy}$ & $\Lambda_{pole}$ & $(\delta_u^{(q)})^{(0)}$
& $\delta_u^{(q)}$ & $\delta_d^{(q)}$ & $\delta_u^{(l)}$ & $\delta_d^{(l)}$
& $\eta^{(q)}$     & $\eta^{(l)}$   \\ \hline
      & $100$ & $(3.6)\cdot10^5$ & $0.0113$ & $0.0090$ & $0.031$ & $0.048$ &
$0.047$
& $0.046$ & $0.039$                 \\ \cline{2-10}
$0.5$ & $150$ & $(1.1)\cdot10^4$ & $0.0093$ &$0.0065$& $0.029$ & $0.061$ &
$0.051$
& $0.044$ & $0.042$                 \\ \cline{2-10}
      & $200$ & $(2.3)\cdot10^3$ & $0.0086$ &$0.0058$& $0.029$ & $0.070$ &
$0.053$
& $0.044$ & $0.044$                 \\ \hline
      & $100$ &$(1.0)\cdot10^{18}$&$0.0095$ &$0.0081$& $0.031$ & $0.030$ &
$0.040$
& $0.047$ & $0.033$                 \\ \cline{2-10}
$1.0$ & $150$ &$(4.1)\cdot10^{10}$&$0.0071$ &$0.0047$& $0.029$ & $0.044$ &
$0.045$
& $0.044$ & $0.038$                 \\ \cline{2-10}
      & $200$ & $(2.5)\cdot10^6$ & $0.0061$ &$0.0035$& $0.027$ & $0.070$ &
$0.053$
& $0.041$ & $0.044$                 \\ \hline
$5.0$ & $200$ &$(1.6)\cdot10^{16}$&$0.0052$ &$0.0028$& $0.027$ & $0.045$ &
$0.046$
& $0.041$ & $0.038$                 \\ \hline
minSM& $200$ &$(2.8)\cdot10^{17}$&$0.0051$ &$0.0029$& $0.059$ & $0.043$ &
$0.138$
& $0.089$ & $0.119$                 \\ \cline{2-10}
      & $250$ & $(1.4)\cdot10^9$ & $0.0045$ &$0.0023$& $0.065$ & $0.077$ &
$0.077$
& $0.100$ & $0.065$                 \\ \hline
  \end{tabular}
 \caption{Quark and lepton flavor democracy (FD) parameters at
$\mu=\Lb_{pole}/2$, and values of $\Lb_{pole}$ (in $GeV$), for various
$v_u/v_d$ (in SM with two Higgs doublets, type II), and for the minimal
SM, and for various $m_t^{phy}$. The Dirac neutrino masses at $\mu = 1 GeV$
were taken to be $m^D_{\nu_{\mu}}= 10 GeV$ and $m_{\nu_{\tau}}=100 GeV$.
$(\delta_u^{(q)})^{(0)}$ is the value of this parameter at $\mu=1 GeV$.
Other values of the FD parameters at $\mu=1 GeV$ are in this case:
$(\delta_d^{(q)})^{(0)}=0.033$; $(\delta_u^{(l)})^{(0)}=0.100$;
$(\delta_d^{(l)})^{(0)}=0.060$;
$(\eta^{(q)})^{(0)}= 0.05$
($(\eta^{(l)})^{(0)}$ is assumed to be $0.05$, for simplicity).
The entries with
$m_t^{phy}=100,150 GeV$ that are missing have $\Lb_{pole}>\Lb_{Planck}$.}
 \end{center}
\end{table}

\begin{table}
\vspace{0.4cm}

\begin{center}
Table 3 \\
\vspace{0.5cm}
  \begin{tabular}{|c|c|c|r|c|} \hline
$v_u/v_d$ & $m_t^{phy}$ & $\Lambda_{pole}$ & $m^{D}_{\nu_{\tau}}$
& $m^{phy}_{\mu_{\tau}}$    \\  \hline
     & $100$ & $(3.20)\cdot10^{13}$ &  $43.04$  &  $(5.80)\cdot10^{-11}$  \\
\cline{2-5}
$0.5$& $150$ & $(3.85)\cdot10^4$    &  $79.02$  &  $(1.62)\cdot10^{-1}$  \\
\cline{2-5}
     & $200$ & $(2.72)\cdot10^3$    &  $93.02$  &  $3.18$          \\
\cline{2-5}
     & $250$ & $(1.18)\cdot10^3$    &  $99.06$  &  $8.28$          \\ \hline
     & $100$ & $>\Lambda_{Planck}$&       &                  \\ \cline{2-5}
$1.0$& $150$ & $(6.78)\cdot10^{16}$ &  $60.28$  &  $(5.36)\cdot10^{-14}$ \\
\cline{2-5}
     & $200$ & $(1.95)\cdot10^6$    & $104.47$  &  $(5.60)\cdot10^{-3}$  \\
\cline{2-5}
     & $250$ & $(2.76)\cdot10^4$    & $127.20$  &  $(5.88)\cdot10^{-1}$  \\
\hline
     & $100$ & $>\Lambda_{Planck}$&       &                  \\ \cline{2-5}
$5.0$& $150$ & $>\Lambda_{Planck}$&       &                  \\ \cline{2-5}
     & $200$ & $(2.50)\cdot10^{20}$ &  $75.22$  &  $(2.26)\cdot10^{-17}$ \\
\cline{2-5}
     & $250$ & $(1.13)\cdot10^8$    & $126.00$  &  $(1.40)\cdot10^{-4}$  \\
\hline
  \end{tabular}
 \caption{Predictions of $\Lb_{pole}$, $m^D_{\nu_{\tau}}$ and
$m^{phy}_{\nu_{\tau}}$ for various values of $v_u/v_d$ and $m_t^{phy}$
(all energies are in $GeV$). For $m^{phy}_{\nu_{\tau}}$
we used the usual see-saw
mechanism with $M_R \approx \Lb_{pole}$ (eqs.~(19),~(20)).}
 \end{center}
\end{table}
\clearpage

\noindent {\Large\bf Figure Captions}

\vspace{0.4cm}

\noindent {\bf Fig.~1:} Quark masses as functions of energy scale $\mu$, in SM
with
two Higgs doublets (type II), for a typical value of VEV ratio $v_u/v_d=1$.
The masses of light quarks ($m_b,m_c,m_s$) and mixings at $\mu=1 GeV$ are
fixed by experiment~\cite{gl1,gl2}, and $m_t^{phy}(=m_t(\mu=m_t)) =
100, 150, 200, 250 GeV$. Leptons were assumed massless.

\vspace{0.3cm}

\noindent {\bf Figs.~2,~3:} Quark flavor democracy parameters $\delta_u^{(q)}$,
$\delta_d^{(q)}$ and $\eta^{(q)}=(V_{ckm}^{(q)})_{cb}$ as functions
of energy scale $\mu$, for VEV ratios $v_u/v_d = 1, \ 0.5$, respectively
(in SM with two Higgs doublets, type II). Other explanations as for Fig.~1.

\vspace{0.3cm}

\noindent {\bf Fig.~4:} Quark flavor democracy parameters $\delta_u^{(q)}$,
$\delta_d^{(q)}$ and $\eta^{(q)}$ in the minimal SM. Other explanations
as in Figs.~1-3.

\vspace{0.3cm}

\noindent {\bf Fig.~5:} The ratio $g_b/g_t$ of Yukawa couplings as a function
of
energy scale, for $m_t^{phy}=200 GeV$ and VEV ratios $v_u/v_d= 0.5$,
$1.0$, $5.0$ (in SM with two Higgs doublets, type II).

\vspace{0.3cm}

\noindent {\bf Fig.~6:} Masses of $\tau$, $b$, $\nu^{Dirac}_{\tau}$ and $t$ as
functions of energy scale $\mu$, for $m_t^{phy}=200 GeV$ and
$v_u/v_d=1$ (in SM with two Higgs doublets, type II). The masses
$m_b$ and $m_{\tau}$ at $\mu = 1 GeV$ are fixed by experiments.
The Dirac $\nu_{\tau}$ masses at $\mu=1 GeV$, which satisfy all the
boundary conditions at $\mu= 1 GeV$ and the high energy boundary conditions
(18), were found by numerical integration of RGE's from $\mu=1 GeV$ to
$\Lb_{pole}$.

\vspace{0.3cm}

\noindent {\bf Fig.~7:} Upper and lower bounds on the values of $m_t^{phy}$ as
functions of $v_u/v_d$ for various specific upper bounds imposed on
$m_t^{phy}$ ($e.~g.~\leq 31 MeV$, or $\leq 1 MeV$, or $ \leq 17 keV$)
and on $\Lb_{pole}$ ($e.~g.~\leq \Lb_{Planck}$, or $\leq 10^{10}GeV$,
or $\leq 10^5 GeV$), respectively.


\newpage

\begin{thebibliography}{99}

\bibitem{dm}
N.G.~Deshpande and E.~Ma, Phys.~Rev.~D18 (1978) 2574;
J.F.~Donoghue and L.-F.~Li, Phys.~Rev.~D19 (1979) 945;
L.J.~Hall and M.B.~Wise, Nucl.~Phys.~B187 (1981) 397

\bibitem{hp}
C.T.~Hill, Phys.~Rev.~D24 (1981) 691;
E.A.~Paschos, Z.~Phys.~C26 (1984) 235;
J.H.~Halley, E.A.~Paschos and H.~Usler, Phys.~Lett.~B155 (1985) 107

\bibitem{gl1}
J.~Gasser and H.~Leutwyler, Phys.~Rep.~87 (1982) 77;
S.~Narison, Phys.~Lett.~B197 (1987) 405

\bibitem{cdf}
CDF Collaboration: F.~Abe et al., Fermilab-Pub-91-352-E (Dec. 1991)

\bibitem{ef}
J.~Ellis and G.L.~Fogli, Phys.~Lett.~B249 (1990) 543;
G.~Altarelli, R.~Barbieri and S.~Jadach, Nucl.~Phys.~B369 (1992) 3;
P.~Langacker, talk given at Madison SSC Workshop (March 1992)

\bibitem{gl2}
Particle Data Group: G.P.~Yost et al., Phys.~Lett.~B239 (1990) 1
(Erratum: Phys. Lett. B253 (1991) 524)

\bibitem{hlr}
C.T.~Hill, C.N.~Leung and S.~Rao, Nucl.~Phys.~B262 (1985) 517

\bibitem{bhp}
V.~Barger, J.L.~Hewett and R.J.N.~Phillips, Phys.~Rev.~D41 (1990) 3421

\bibitem{mv}
M.E.~Machacek and M.T.~Vaughn, Nucl.~Phys.~B249 (1985) 70;
M.~Fischer and J.~Oliensis, Phys.~Lett.~B119 (1982) 385;
D.R.T.~Jones, Phys.~Rev.~D25 (1982) 581
I.~Jack and H.~Osborn, Nucl.~Phys.~B249 (1985) 472

\bibitem{grs}
T.~Yanagida, Proceedings of the Workshop on Unified Theory and Baryon
3Number of the Universe (KEK, Japan, 1979);
M.~Gell-Mann, P.~Ramond and R.~Slansky, in ``Supergravity'', edited
by P.~Van Nieuwenhuizen and D.Z.~Freedman (North-Holland, Amsterdam, 1979)

\bibitem{argus}
ARGUS Collaboration: H.~Albrecht et al., Phys.~Lett.~B202 (1988) 149,
B292 (1992) 221;
T.K.~Kuo and J.~Pantaleone, Rev.~of Mod.~Phys., Vol.~61, No.~4 (1989) 937

\bibitem{hpr}
A.~Hime, R.J.N.~Phillips, G.G.~Ross and S.~Sankar, Phys.~Lett.~B260
(1991) 381

\bibitem{h2condens}
M.~Suzuki, Phys.Rev.~D41 (1990) 3457; M.~Harada and N.~Kitazawa,
Phys.~Lett.~B257 (1991) 383

\bibitem{bo}
R.~B\"onisch, Munich preprint LMU-91/03 and private communication

\bibitem{svz}
M.A.~Shifman, A.I.~Vainshtein and V.I.~Zakharov, Nucl.~Phys.~B120 (1977)
316

\bibitem{eg}
D.~Bailin and A.~Love, ``Introduction to Gauge Field Theory''
(Adam Hilger, Bristol and Boston, 1986);
T.~Eguchi, Phys.~Rev.~D14 (1976) 2755

\bibitem{zb}
Z.~Berezhiani, private communication

\end{thebibliography}
\end{document}